**Uncovering the role of flow rate in redox-active polymer flow batteries: simulation of reaction distributions with simultaneous mixing in tanks**

V. Pavan Nemani[a,d] and Kyle C. Smith[a,b,c,d]*

[a]Department of Mechanical Science and Engineering, [b]Computational Science and Engineering, [c]Beckman Institute, University of Illinois at Urbana-Champaign, Urbana, IL, USA and [d]Joint Center for Energy Storage Research, USA

*corresponding author's email: kcsmith@illinois.edu

**Abstract**

Redox flow batteries (RFBs) are potential solutions for grid-scale energy storage, and deeper understanding of the effect of flow rate on RFB performance is needed to develop efficient, low-cost designs. In this study we highlight the importance of modeling tanks, which can limit the charge/discharge capacity of redox-active polymer (RAP) based RFBs. The losses due to tank mixing dominate over the polarization-induced capacity losses that arise due to resistive processes in the reactor. A porous electrode model is used to separate these effects by predicting the time variation of active species concentration in electrodes and tanks. A simple transient model based on species conservation laws developed in this study reveals that charge utilization and polarization are affected by two dimensionless numbers quantifying (1) flow rate relative to stoichiometric flow and (2) size of flow battery tanks relative to the reactor. The RFB's utilization is shown to increase monotonically with flow rate, reaching 90% of the theoretical value only when flow rate exceeds twenty-fold of the stoichiometric value. We also identify polarization due to irreversibilities inherent to RFB architecture as a result of tank mixing and current distribution internal to the reactor, and this polarization dominates over that resulting from ohmic resistances particularly when cycling RFBs at low flow rates and currents. These findings are summarized in a map of utilization and polarization that can be used to select adequate flow rate for a given tank size.

# 1. Introduction

The transition from non-renewable to intermittent renewable energy sources necessitates the development of technologies for grid-scale energy storage systems [1]. Redox flow batteries (RFBs) are one technology that promises independent control over energy capacity (system size) and power density (reactor design), which is a property that is ideal for grid-scale applications. The long lifetime of RFBs [2–4] relative to other technologies is also a distinct advantage for various grid uses where capital investment is significant.

A variety of active compounds, inorganic (e.g. vanadium [5], polysulfide-bromine [6]) and organic/semi-organic (e.g. TEMPO [7], quinone [8], viologen based polymers [9]) have been explored for use in RFBs. Vanadium redox batteries [5] are the most common and widely studied aqueous redox flow batteries (AqRFBs). However, metal ions such as vanadium are not abundantly available and their high cost restricts their utility for large scale systems [2,3]. The corrosive and toxic nature of some of the non-metallic materials, such as bromine, affects the pipes, pumps and tanks [10]. This motivated the development of AqRFBs with organic or semi-organic active species [8,11]. Such systems can be cheaper and safer, but the electrochemical stability window of aqueous based electrolytes impedes achievement of high energy density. On the other hand, nonaqueous redox flow batteries (NAqRFBs) provide a wider electrochemical stability window (hence high energy density) [12–15], while NAqRFBs tend to have lower ionic conductivity and higher electrolyte costs as compared to their aqueous counterparts [16].

Irrespective of the type of solvent employed (i.e., aqueous or non-aqueous), the crossover of active species is a degradative mechanism that must be managed to make efficient, long-life RFBs. Much research has been conducted to address this by the development of ion-selective membranes that inhibit crossover by electrostatic and steric effects, particularly in the context of vanadium-based RFBs [17,18]. An alternative approach to preventing crossover is to engineer active species with macromolecular motifs that can be excluded by porous separators. Drawing on this idea organic redox-active moieties bound to a polymer backbone were synthesized previously, giving rise to redox-active polymers (RAPs) dissolved in non-aqueous solvent [9]. Here, polymers ranging between 21 kDa to 318 kDa molecular weight were found to have a range of permeabilities through Celgard separators, with the largest polymers being rejected at a rate of 86% relative to the small-molecule supporting salt species also in solution [9]. Later similar approaches were used to make aqueous RAP electrolytes [11]. Following this work, other macromolecular motifs have been developed, including redox-active colloids [19].

Solutions containing macromolecular motifs, such as RAPs, show high viscosity with increasing redox-active species concentration, [9] potentially affecting mechanical design and pumping costs in scaled RFBs. Fluid distribution within the reactor, and consequently reaction distributions, can be affected by the electrolyte viscosity, in addition to the flow rate used. Therefore, understanding the impact of flow rate on cell performance is essential to establish optimized operating regimes and cell designs for RAP-based RFBs. Tailoring flow field design (serpentine, parallel, interdigitated and spiral flows) and electrode architecture is important to enhance mass transport in the reactor, thus achieving high power densities and limiting currents [20–22]. Systematic studies in this regard revealed that an interdigitated flow field (IDFF) paired with carbon paper electrodes gives the optimal performance with respect to electrochemical reactions and pressure drops [22]. Therefore, proper reactor design coupled with material selection [14,16,23] would significantly reduce the cost of the RFBs helping in their widespread commercialization [24]. Also, less conventional RFBs utilizing high-viscosity, shear-thinning suspensions of carbon black and solid active compounds required the development of various flow-mode strategies to enable efficient operation in both aqueous and non-aqueous RFBs [25–29].

Simulation of local electrochemical processes within RFBs provides mechanistic insight into the system dynamics of RFBs that can be used to develop design and operational guidelines. Various models that predict the performance of RFBs have been developed in the past, most of which are focused around vanadium RFBs [30–36]. In addition, models for polysulfide redox flow battery [26,37], hydrogen bromine flow battery [38,39] and metal free organic-inorganic flow batteries [40] have been developed. However, most of these models are focused primarily on the reactor, assuming steady or pseudo-steady state conditions (e.g., constant concentration input to the reactor) and were specific to particular chemistries. Therefore, models that simulate the reactor and tanks in a coupled manner, and under transient conditions, are needed to understand the impact of flow rate on capacity loss in RFBs.

In the present work, we simulate the cycling of RAP-based RFBs using two-dimensional porous electrode theory [41,42]. Our approach consists of modeling tanks as well-mixed units, and we show that the transient response of the redox-active species concentration in the tank at various flow rates is the dominant factor affecting the charge utilization in these RFBs. Further, with the help of a simplified model, we identify two dimensionless numbers (one for tank size and another for flow rate) that determine the utilization of charge capacity and polarization within these RFBs. We also identify certain inherent irreversibilities causing polarization apart from the ohmic resistances of the RFBs. We use these results to determine the operating regimes of flow that are needed to achieve certain utilization for the entire system.

## 2. Modeling Approach

In the present work we model RFBs using redox-active polymer (RAP) solutions as electrolytes. Porous electrode theory is used to obtain the transient distribution of active species concentration by incorporating microscale processes (solution advection, ion conduction, electron conduction, and reaction kinetics). The effect of mixing within tanks is captured by coupling the concentrations of active species entering and leaving the reactor.

### 2.1 Configuration of reactor and tanks

The RFBs simulated here (Fig. 1a) consist of a reactor and two tanks which store the electrolyte. Electrochemical reactions occur in the two electrodes of the reactor that are separated by a membrane and are sandwiched between two electron-conducting plates (collectors). Carbon felt is used as the electrode material due to its high porosity and large specific area per unit volume. Because RAPs are known to prevent crossover due to size exclusion [9], we model a porous polymer separator in place of a membrane, and we assume negligible crossover.

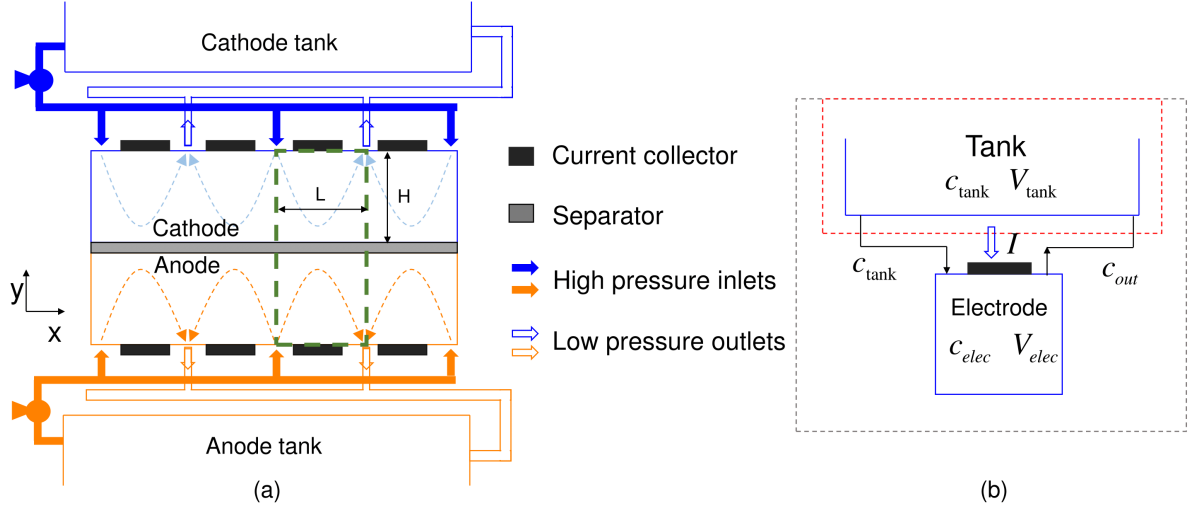

**Figure 1:** (a) Schematic of the simulated flow battery using a 2D interdigitated flow field (IDFF) in its reactor. *H* is the electrode thickness and *L* is the length of representative basic repeat unit (green box) modeled here. (b) Diagram of one half of the RFB unit cell with two control volumes.

The specific reactor design that we employ uses an interdigitated flow field (IDFF) that has produced satisfactory electrochemical performance and pressure drop in vanadium-based RFBs [21,22]. Assuming sufficiently small pressure drop within IDFF channels [43] (assuming low aspect ratio channels), we model the reactor as a two-dimensional domain consisting of periodically repeating units of length $L$ (Fig. 1a, green dashed box). Each electrode with open pore volume $V_{elec}$ is supplied with RAP solution at a volumetric flow rate $\dot{V}$ from a tank of certain volume $V_{tank}$. RAP solution exiting the reactor is transported back to the tank, where it is assumed to mix *perfectly* with solution stored in the tank.

*2.2 Electrochemical kinetics and transport processes*

A transient two-dimensional model based on porous-electrode theory [42] is developed to obtain reaction distributions by incorporating simultaneous ion conduction, electron conduction, and electrochemical kinetics. With this approach macroscopic transport processes within the electrodes are modeled using a volume-averaged approach, while reactions occur at the scale of microscopic pores within the carbon felt.

Presently we assume the electrochemical kinetics of RAP solutions to be consistent with the behavior for individual monomer units. Because the polymeric motif [9] can, in theory, be applied to arbitrary active molecules, we consider the properties of a specific RAP (benzyl ethyl viologen) as a surrogate for generic RAP molecules. We assume that, in both electrodes, one electron is released during oxidation of the reduced species $R$ to the oxidized species $O$ (i.e., $R^{z_R} \rightarrow O^{z_O} + e^-$). We model electrochemical kinetics using the Butler-Volmer equation for the pore-scale reaction current-density $i_n$ [44]

$$i_n = i_0 \left[ \exp\left(\frac{0.5 F \eta}{RT}\right) - \exp\left(-\frac{0.5 F \eta}{RT}\right) \right]. \tag{1}$$

The driving force for this electrochemical reaction is the overpotential at the electrode/electrolyte interface $\eta = \phi_s - \phi_e - \phi_{eq}$, where $\phi_{eq}$ is the equilibrium potential of the active compound that depends on the state of charge/discharge and is determined from the Nernst equation as follows:

$$\phi_{eq} = E^0 - \frac{RT}{n_e F} \ln \frac{c_R}{c_O}, \qquad (2)$$

where $E^0$ is the equilibrium potential at 50% state of charge, $c_R$ and $c_O$ are the concentrations of the active species in reduced and oxidized state respectively, $T$ is the temperature (298K) and $F$ is Faraday's constant. Here $n_e = 1$ for a one-electron transfer redox process.

The exchange-current density $i_0$ is related to the reaction rate-constant $k$ as:

$$i_0 = Fk(c_O)^{0.5}(c_R)^{0.5}. \qquad (3)$$

The transport of ions through the electrode's thickness is coupled to electrochemical kinetics using a volume-averaged approach, where the locally averaged pore-scale concentration $c_i$ of each species $i$ is governed by mass conservation:

$$\varepsilon \frac{\partial c_i}{\partial t} + \nabla \cdot (\vec{N}_i) + S_i = 0, \qquad (4)$$

where $\vec{N}_i$ is superficial flux at the macro-scale, and $S_i$ accounts for the generation and consumption of species $i$ by electrochemical reactions. Here, for the one-electron transfer reactions that we consider $S_i$ is related to the pore-scale reaction current-density as $S_R = ai_n/F$ and $S_O = -ai_n/F$ for reduced and oxidized species, respectively. The surface area per unit electrode volume of the electrode $a$ is determined by assuming the carbon fibers in the electrode to be cylindrical with 10μm diameter and by setting porosity of the electrode $\varepsilon$ to 90% (see Table 1).

**Table 1**: Reactor geometry and carbon felt electrode properties

| Parameter | Value |
|---|---|
| length of current collector, $L(mm)$ | 2 |
| electrode thickness, $H(\mu m)$ | 200 |
| electrode flow entry length, $L_{ent}(mm)$ | 0.5 |
| porosity of carbon felt, $\varepsilon$ | 0.9 |
| solid volume fraction of carbon felt, $v_s$ | 0.1 |
| permeability of electrolyte in carbon felt, K (m$^2$) | 6x10$^{-11}$ (Ref.[30]) |

In dilute-solution theory the Nernst-Planck equation accounts for diffusion (with effective diffusion coefficient $D_i^{eff}$), migration, and convective fluxes [44,45]. The total flux for a superficial velocity profile $\vec{u}$ in the electrode region is given as:

$$\vec{N}_i = -D_i^{eff}\nabla c_i - \frac{z_i c_i D_i^{eff}}{RT}F\nabla\phi_e + \vec{u}c_i. \tag{5}$$

Presently, we assume an excess of supporting electrolyte, such that the effects of migration on RAP molecules can be neglected [44]. Also, for the conditions of interest in the present work, we find that the Peclet number (defined as $Pe = |\vec{u}|L/D^{eff}$, representing the characteristic advection transport relative to that of diffusion) is greater than 100 (for viologen-based RAPs [9]) for the conditions of interest in this work. As such, we neglect the effect of diffusive flux through the electrode thickness. Employing these approximations, the conservation equation for the reduced species in either electrode simplifies to:

$$\varepsilon c^0 \frac{\partial \psi}{\partial t} + c^0 \nabla \cdot (\vec{u}\psi) + a\frac{i_n}{F} = 0, \tag{6}$$

where $\psi$ is the state of discharge, defined as $\psi = \frac{c_R}{c^0}$ where $c_R$ is the concentration of the redox-active species in its reduced state and $c^0$ is the initial tank concentration.

Ion conduction within the supporting electrolyte is modeled here assuming uniform concentration of the inactive, conductive species, such that diffusion currents are negligible and the solution's effective ionic conductivity $\kappa_{eff}$ is a constant (a common assumption in potential theory [44]). Following from these assumptions solution-phase current conservation takes the form:

$$\nabla \cdot (-\kappa_{eff}\nabla\phi_e) - a\nu_s i_n = 0. \tag{7}$$

Kinetics also couple to the solid-phase potential $\phi_s$ within the carbon felt through electronic current conservation:

$$\nabla \cdot (-\sigma_s \nabla \phi_s) + a\nu_s i_n = 0, \tag{8}$$

where $\sigma_s$ is effective electronic conductivity.

The aforementioned processes occur simultaneously as solution flows through each electrode. The corresponding superficial velocity field $\vec{u}$ results from solution viscosity $\mu$, gradients of pressure $p$, and permeability of the carbon felt $K$, based on Darcy's law, $\vec{u} = -(K/\mu)\nabla p$, and mass conservation, $\nabla \cdot \vec{u} = 0$ for a constant electrolyte density.

The system properties used in this simulation model are listed in Table 1. Both catholyte and anolyte have the same properties (Table 2) except for the equilibrium potentials of the redox reactions in the cathode and anode. The separator is modeled as a porous medium with certain porosity and tortuosity (see Table 3) which affects the ion conduction through the membrane.

**Table 2:** Electrochemical-transport material-parameters of the carbon felt electrode and redox active electrolyte used in the present simulations

| Parameter | Anode | Cathode |
|---|---|---|
| electronic conductivity of electrode, $\sigma_s$ (S/m) | 100 | 100 |
| volumetric surface-area of carbon felt, $a/v_s$ (m$^2$/m$^3$) | 4×10$^5$ | 4×10$^5$ |
| rate constant for the redox reactions, $k$ (m/s) | 3 x10$^{-5}$[‡] | 3x10$^{-5}$ [‡] |
| equilibrium potential of the redox species at 50%SOC (V) | 0 | 3 |
| ionic conductivity of the electrolyte, $\kappa_0$ | 1.58 (0.1M) 2.31 (0.5M) | 1.58 (0.1M) 2.31(0.5M) |

[‡]Ref. [54]

**Table 3**: Ion conduction properties of separator (Ref. 54).

| Parameter | Value |
|---|---|
| thickness, $H_{sep}$ ($\mu$m) | 25 |
| permeability, K (m$^2$) | 1x10$^{-12}$ |
| porosity, $\varepsilon_{sep}$ | 0.3 |
| tortuosity, $\tau_{sep}$ | 6 |

The reactor is subjected to galvanostatic cycling where a constant current density, $i_{app}$, is applied to the cathode current collector. The separator is electronically insulating. Symmetric boundary conditions are used for the solution-phase potentials, solid-phase potentials and redox-active concentration on the left and right boundaries of the 2D unit cell. The superficial velocity of the redox-active species into the electrode region is fixed for each particular case simulated here. A particular flow rate sets the superficial velocity entering the electrode region through channel openings of length $L_{ent}$, and this is used to determine the 2D flow velocities from Darcy's law.

The governing equations are discretized using the finite-volume method with implicit differencing in time and central differencing in space [46]. This scheme was implemented in MATLAB using an iterative algorithm, where the coupled set of governing equations is solved iteratively to resolve non-linearities [28]. At every iteration the algebraic-multigrid method is used to solve the discrete, linearized equations assembled in a sparse matrix [47–50]. Convergence is achieved when potentials differ by less than 10$^{-8}$ V and redox-active species concentrations differ by less than 10$^{-8}$ mol/L during successive iterations.

The numerical implementation of the governing equations was verified with analytical solutions under certain conditions. To verify species conservation equations a test was performed with twenty times the stoichiometric flow rate with zero applied current, in which case the total amount of each species in the RFB was conserved to within machine precision. The advection of active species was verified by testing a 20 cm reactor, so as to approach a one-dimensional superficial velocity field. To this reactor a discontinuity in concentration was introduced at the inlet, and the discontinuity was found to propagate at a velocity within 0.1% of the value expected based on the applied flow rate that is attributable to numerical diffusion. Lastly, the RFB with no flow was operated at large Wagner number (representing the scale of kinetic overpotential relative to ohmic polarization and defined as $Wa = \dfrac{\kappa_{\mathit{eff}} RT}{k v_s a c^0 H^2 F^2}$), such that kinetic polarization dominates and active species concentration is uniform. In practice, ionic conductivity was increased two-fold and the reaction rate constant was decreased by two orders of magnitude with respect to the properties listed in Table 2. We find that for a constant applied current density of 1.0 mA/cm$^2$, the concentrations of active species are uniform to within 1% of $c^0$. Also, the value of the reaction current density, as determined by current balance over the current collector and carbon felt, agrees with simulated values within 1%, confirming the implementation of the kinetic model.

## 3. Results and Discussion

The primary aim of this study is to elucidate the role of flow rate in determining the energy-storage performance of RAP-based RFBs. To do this we first explore the galvanostatic cycling characteristics of an RFB for low and high flow rates. In all simulations we choose a target charge/discharge time of 5 hours, which is typical for grid-scale energy storage applications. With such a constraint on cycling, the current density applied to each cell that is simulated will depend on the size of the RFB's tanks relative to the electrolyte volume within the reactor. We start by considering small systems for which the current density is also small (1 mA/cm$^2$). Later, we confirm that the findings obtained under such conditions are similar to those obtained for large systems with higher current densities.

Though polarization decreases with increasing flow rate (as a result of the increased uniformity of current density across the separator), we find that the polarization due to transport processes within the electrodes is not the dominant factor responsible for capacity loss when cycling at low flow rates. Subsequently, we examine the temporal dynamics of active species concentration within the tanks, and find, instead, that capacity loss under such conditions is primarily a result of mixing between solution entering the RFB's tanks and solution stored in the tanks themselves. Subsequently, we present a simplified model for charge utilization and use it to develop maps of charge utilization and polarization as a function of non-dimensional flow rate and tank size.

We define several non-dimensional parameters that we use to correlate results. The utilization $\chi$ is defined as the fraction of theoretical capacity utilized during either charge or discharge of a given cycle, where theoretical capacity $Q_{theory}$ is expressed as $Q_{theory} = (V_{tank} + V_{elec})c^0 F$. We also use non-dimensional parameters to quantify the size of and the flow rate within an RFB: the tank-to-electrode volume ratio, defined as $\alpha = V_{tank}/V_{elec}$, and the non-dimensional flow rate, $\beta = \dot{V}/\dot{V}_{stoich}$, respectively. Here, the stoichiometric flow rate $\dot{V}_{stoich}$ is set by the total solution volume within a given half of the RFB divided by the theoretical charge time $\tau_c$, $\dot{V}_{stoich} = (V_{tank} + V_{elec})/\tau_c$, where $\tau_c = Q_{theory}/I_{applied}$ for a total current $I_{applied}$ at the current collectors. The polarization $\Delta\phi$ is calculated based on simulated results as half the difference of average voltages during charge and discharge.

We use galvanostatic conditions to charge and discharge for each case that follows, so as to produce a theoretical charge/discharge time of 5 hours in each case. In most cases considered the corresponding current density is 1 mA/cm$^2$, and we show that consistent behavior is observed at current densities as high as 10 mA/cm$^2$. Also, charge and discharge processes are terminated at cell voltages of 3.35V and 2.65V, respectively. Figure 2 shows the variation of cell voltage with time for the first five charge/discharge cycles with a particular actives concentration of $c^0 = 0.5M$, tank-to-electrode volume ratio $\alpha = 20$, and non-dimensional flow rate $\beta = 20$. Because the catholyte is oxidized and the anolyte is reduced during charging, cell voltage increases until it reaches the upper voltage limit. This particular case shows near-theoretical utilization during the first charge step, but charge and discharge capacity decrease and approach an asymptote after a certain number of cycles. We refer to a cycle exhibiting this asymptotic behavior as a "limit cycle." Subsequently, we present performance metrics for limit cycles to avoid anomalous capacity variations due to short-time behavior. In practice, we obtain metrics for limit cycles by simulating a finite number of cycles and classify a limit cycle as a cycle with coulombic efficiency in excess of 99.8%.

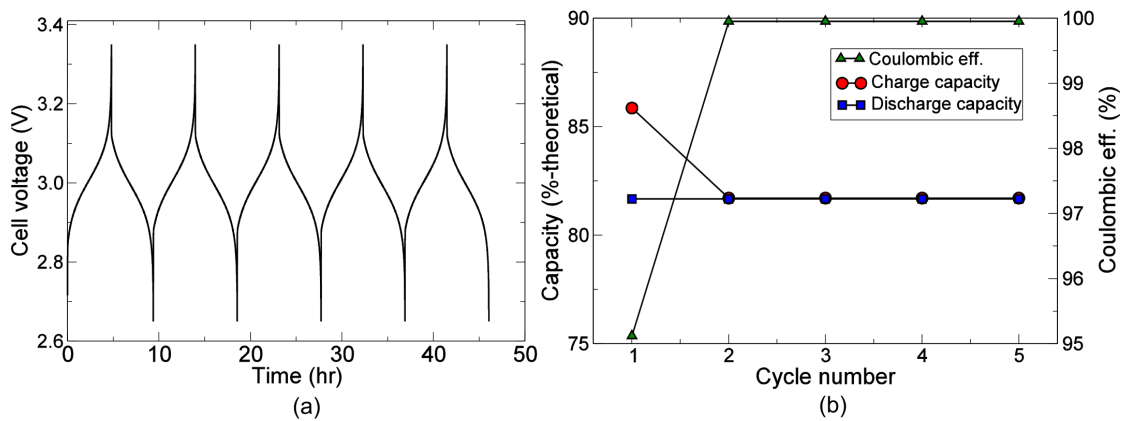

**Figure 2:** (a) Variation of cell voltage with time and (b) variation of charge/discharge capacities and coulombic efficiency with cycle number for an RFB configuration having an initial active-species concentration $c^0$ = 0.5M having a tank-to-electrode volume ratio $\alpha$ = 20 operated at $\beta = 20$.

*3.1 Cycling behavior at flow rate extremes*

We first explore the effect of flow rate on cycling performance by considering two extreme values of flow rate. Here, we consider multiples of the stoichiometric flow rate, because, in theory, the stoichiometric flow rate is the smallest flow rate that can be maintained without consuming an electrolyte's charge capacity prior to exiting the electrode. Specifically, we compare the performance of cases with $\beta=2$ and with $\beta=20$, using twice and twenty times the stoichiometric flow rate, respectively. The limit-cycle cell-voltage curves for these two cases are shown in Fig. 3, as a function of the cell's average state-of-charge. On these plots the terminal state-of-charge determines utilization and polarization is the difference between the charge and discharge voltage curves. At low flow rates, only 20% utilization is achieved, whereas at high flow rates, utilization is approximately 90%. The RFB using low flow rate experiences higher polarization (110mV) than at high flow rate (11mV). Experimentally, similar results have been observed [22] with respect to the voltage curves (better capacity) and energy capabilities (high peak power density and limiting current). They hypothesized that high flow rates would enable greater utilization of the electrode volume, thus leading to better performance. Also previous vanadium RFB models have predicted improved coulombic efficiency with increased flow rate [30].

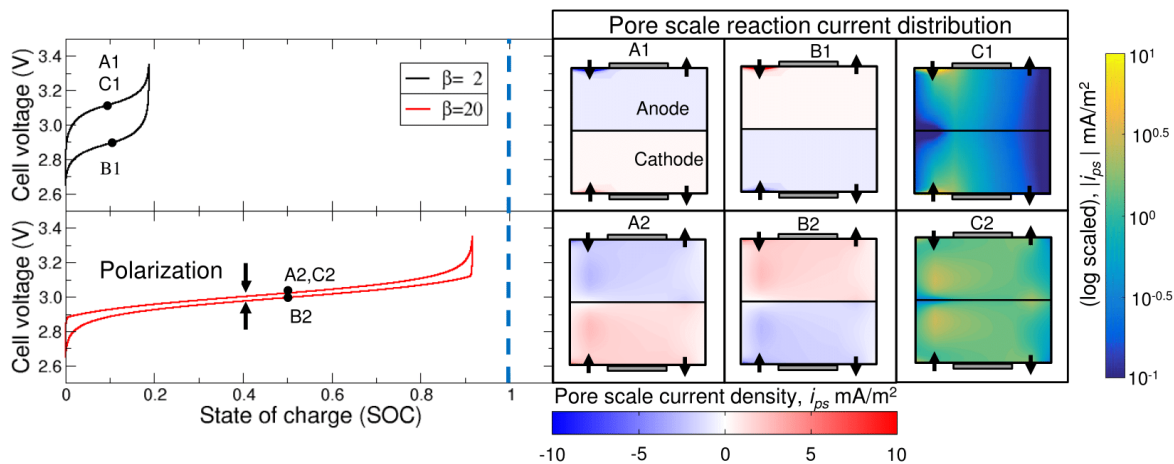

**Figure 3:** Cell voltage variation with state of charge (SOC) during charge and discharge process at low ($\beta=2$) and high ($\beta=20$) flow rates for a RFB configuration having an active species concentration $c^0$ = 0.5M and tank-to-electrode volume ratio $\alpha$ = 20. Snapshots of the pore scale reaction current density with linear and logarithmic contour mapping (geometry not drawn to scale) halfway through the charge and discharge process are shown along with flow directions.

To link the macroscopic trends of polarization with flow rate the spatial variation of the pore-scale reaction current density $i_n(x,y,t)$ are shown in Fig. 3 for times half-way through state-of-charge range accessed during a limit cycle.  This current density is generated at the microscopic interface between carbon-felt fibers and electrolyte (see Sec. 2 and Eq. 1), and, consequently, it is substantially smaller than the average electronic current density applied to the reactor, due to the carbon felt's high specific surface area.  We note that the distributions at the specific instants in time chosen are similar to those observed over the entire charge/discharge process.  These distributions reveal that reaction current density is concentrated at the entry region for both anode and cathode at low flow rate, while the remainder of the electrode is practically inactive (A1 and B1).  Regions of high reaction rates (also called "hot spots") and low reaction rates (also called "dead zones") are made even more apparent by examining reaction current distribution on a logarithmic scale (C1 and C2).  When flow rate is slow enough hot spots form at the electrode entry, while at high flow rates, the hot spot migrates closer to the separator in both electrodes.  In general, reaction rates are more uniformly distributed along the separator at high flow rate than at low flow rate.  Thus at high flow rates, the electrode volume is efficiently utilized, eliminating dead zones across the electrode's length.  These observations also help to explain the reduced polarization observed at high flow rates; the path length through which ions must transport decreases when reactions shift closer to the separator at high flow rate. In contrast, the hotspots at low flow rates are farther away than at high flow rates, leading to high polarization. This reaction localization effect can also be observed in the solution-phase current density across the separator, as shown in Fig. 4.  At low flow rates, the solution-phase current density is maximized near inlets.  As flow rate increases this current distribution becomes more uniform, enabling more efficient transport of ions between electrodes.

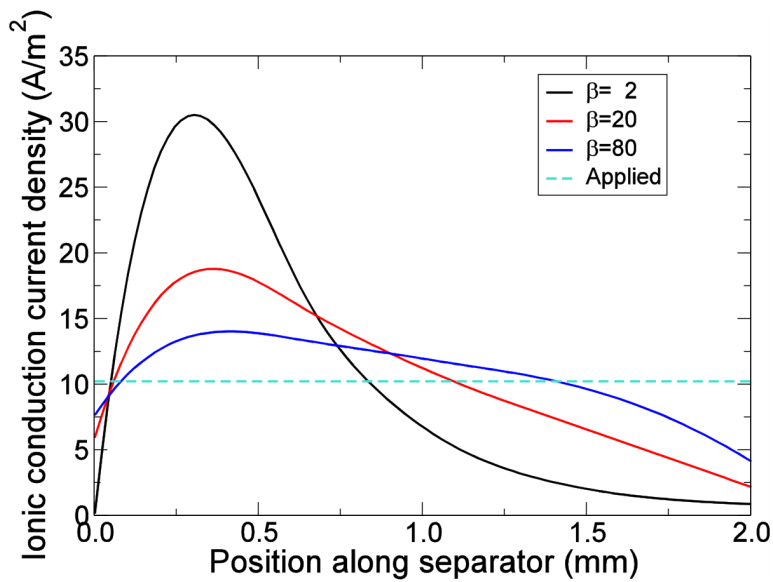

**Figure 4:** Variation of thru-plane (x-direction) ionic current density along the separator for different flow rates in a RFB with initial ion concentration $c^0$ = 0.5M and tank-to-electrode volume ratio $\alpha$ = 20. The dashed line shows the electronic current density applied to the current collectors.

The variations of reaction-rate distributions observed with flow rates can explain the trends of polarization, but polarization alone is not the root cause of the low capacity obtained at low flow rates. In general, when electrochemical cells (including flow batteries and Li-ion batteries) are cycled at high enough rate, polarization can reduce utilization as a result of the finite size of the voltage window over which cycling occurs. For the lowest flow rate investigated thus far ($\beta$=2) polarization is approximately 110 mV, which is substantially smaller than the 700 mV range through which the RFB was cycled (2.65V to 3.35V). Furthermore, the sharp rise and fall in potential at the termination charge and discharge, respectively, suggests that active-species capacity has been locally exhausted, rather than due to the effect of polarization.

*3.2 Effect of tank mixing on capacity loss*

Investigation of mechanisms other than reactor polarization is required to identify the dominant sources of capacity loss at low flow rates and to identify operating conditions to achieve satisfactory utilization during RFB operation. Though we are unaware of prior studies focused on the effects of tanks in conventional RFBs, previous work on suspension-based flow batteries showed that dispersive mixing of charge [28,51] and extension of electrochemical reactions outside of the reaction zone defined by metal current collectors [29] can result in the loss of capacity and energy efficiency during a complete charge/discharge cycle. Consequently, we explore the effect of mixing processes within the RFB's tank on the capacity loss observed when operating near stoichiometric conditions.

Here, we find that capacity loss near stoichiometric conditions arises primarily due to mixing within tanks, and we first illustrate this phenomenon with the aid of an idealized batch-mode operating scheme for an RFB. With this operating scheme solution is pumped into the electrode in discrete batches and the residence time of each batch in the reactor sets the time-averaged flow rate through the reactor. We note that similar schemes, referred to as the "intermittent flow mode," have been employed with suspension-based RFBs to enable efficient operation [26,28,51]. Figure 5 shows half of two batch-mode RFBs: one is operated at low time-averaged flow rate ($\beta=1$) and the other at high flow rate ($\beta=10$). For this example, tanks are chosen to be twice as large as the electrode ($\alpha=2$) and to contain electrolyte with 100% state-of-discharge at the beginning of the first charging cycle.

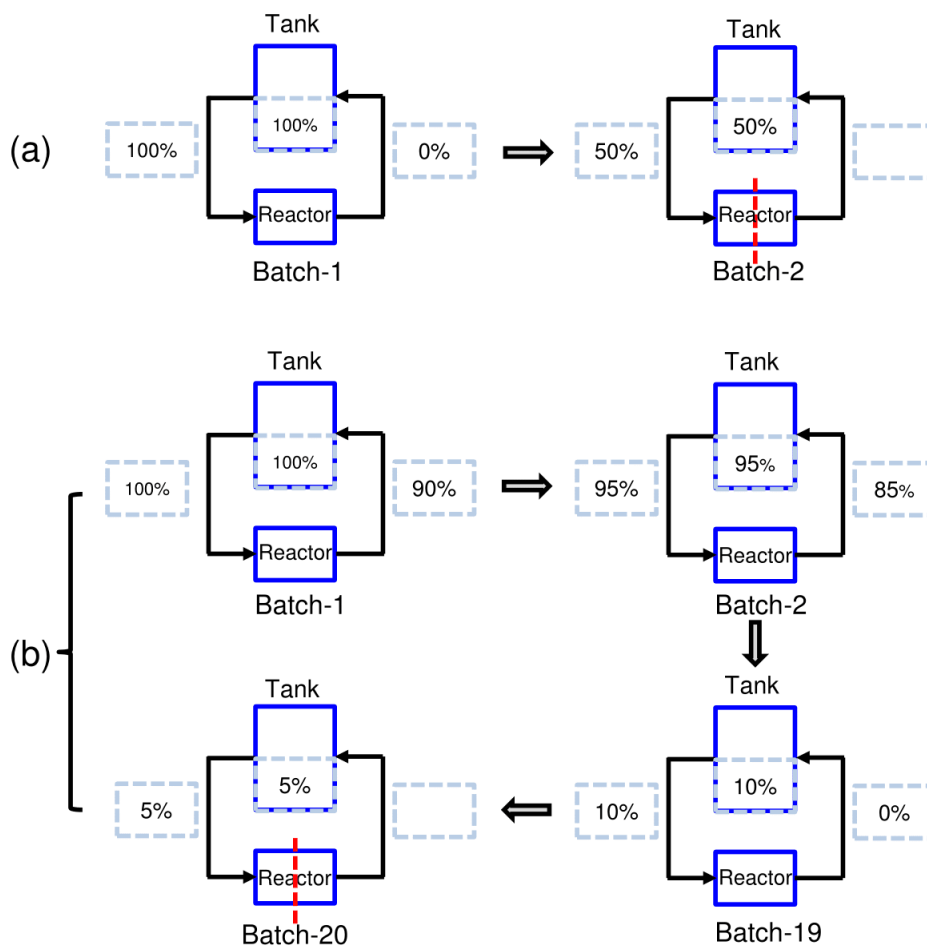

**Figure 5:** Schematics depicting variation of state-of-discharge in a batch mode operated RFB with $\alpha=2$ operated at (a) stoichiometric flow rate and at (b) ten times the stoichiometric flow rate.

Figure 5a shows the corresponding state-of-discharge variation for a stoichiometric flow rate ($\beta=1$) within the tank and within the inlet and outlet of the electrode. The residence time for stoichiometric conditions we denote as $\tau_{stoich}$. By definition, under stoichiometric conditions, the electrolyte's capacity is consumed after a single pass through the electrode by cycling for a time period of $\tau_{stoich}$. In this particular representation, one batch of solution (equal to the reactor volume) enters the reactor at 100% state-of-discharge and exits the electrode at 0% state-of-discharge. Subsequently this electrolyte, which is fully charged, is now combined with solution in the tank, where it is mixed to form solution with 50% state-of-discharge. The next batch of solution entering the electrode is at 50% state-of-discharge, which contains substantially less charge capacity than the first batch. As a result, this solution's capacity will be exhausted in a time period 50% shorter than that of the first batch. As a result, the total charge time for two batches is approximately $1.5\tau_{stoich}$ with stoichiometric flow, while the cell should have charged for $3.0\tau_{stoich}$ if mixing had not occurred within its tanks. Thus, this cell produces a utilization of approximately 50%. The corresponding capacity loss is further evidenced by the remaining solution within the RFB's tank, which is filled with solution having 50% state-of-discharge. With larger, more practical tank-to-electrode volume ratios, capacity losses due to mixing are even more significant.

Operating the RFB at high flow rates can improve charge utilization by minimizing the difference in state-of-discharge between the tank and electrode. Figure 5b shows an RFB operated in batches with ten times the stoichiometric flow rate ($\beta=10$) for the same tank size as in the previous example. At this flow rate, residence time reduces tenfold to $0.1\tau_{stoich}$. Therefore, the state-of-discharge consumed during a single pass of each batch is 10%, and the state-of-discharge for the first batch reduces from 100% to 90% (Fig. 5b,i) when the batch is finished. When combined with solution in the tank its state-of-discharge decreases to 95% (Fig. 5b,ii). The next batch enters with 95% state-of-discharge. After 19 such batches the tank concentration reduces to 5% (Fig. 5b,iv). As a result, the batch's capacity is exhausted after $0.05\tau_{stoich}$ during the 20th batch, resulting in 95% utilization.

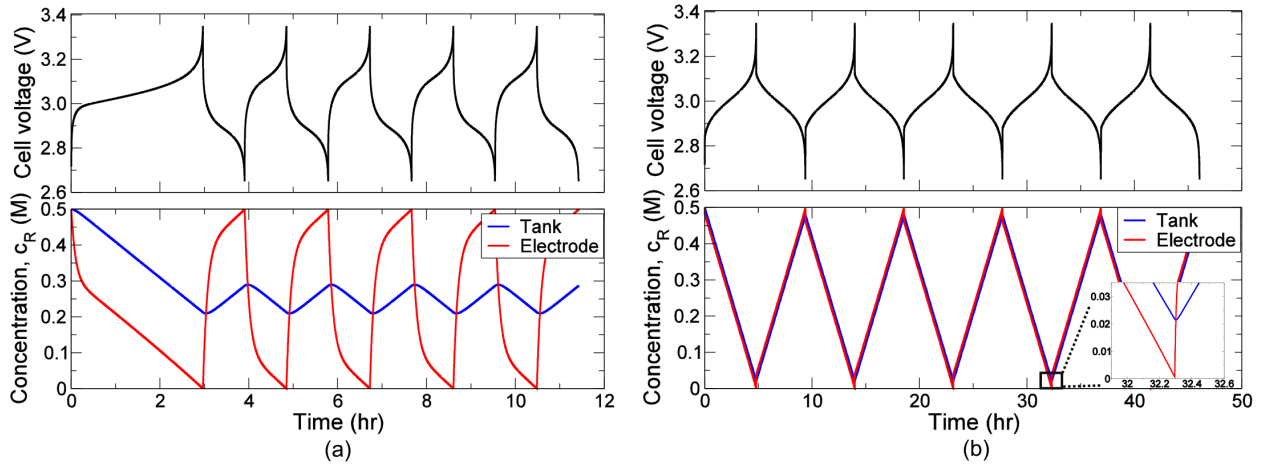

**Figure 6:** Variation of reduced active species concentration in tank and electrode (space averaged) for (a) low ($\beta=2$) and (b) high ($\beta=20$) flow rates in an RFB with initial active-species concentration $c^0$ = 0.5M and tank-to-electrode volume ratio $\alpha$ = 20.

Similar dynamics of actives concentration within tanks and electrodes are observed when using continuous flow coupled to electrode reactions. Figure 6 shows the average concentration of reduced species in the cathode tank and electrode, in addition to the corresponding cell voltage profiles for a RFB with an initial tank concentration $c^0 = 0.5M$ and $\alpha = 20$ at two different flow rates. At low flow rate during the first charging step (Fig. 6a), the average concentration in the electrode rapidly drops to around $c^0/2 = 0.25M$. Thereafter, concentrations in the tank and electrode evolve at similar rates until concentration within the electrode reduces to zero. At the same instant in time, cell voltage diverges due to the exhaustion of capacity within the electrode, while the tank concentration is approximately 0.24M. When a subsequent discharge step begins immediately following the end of the charge step, active species in the electrode are initially in a fully reduced state, producing a rise in reduced species concentration with time when solution from the tank is pumped to it with a concentration of 0.24M. After sufficient time discharging, reduced species concentration in the electrode increases to 0.5M and no oxidized species exist. Consequently, capacity is exhausted at this instant in time. As a result of the short duration of the discharge process, concentration within the tank increases by a small degree during the discharge process (from 0.24M to about 0.26M). The next charge cycle starts with the tank now at approximately 0.26M. This process repeats as a limit cycle for the remainder of charge/discharge cycles, where tank concentration fluctuates between 0.24M and 0.26M, and produces a miniscule charge utilization of approximately 4%.

In contrast, when flow rate is high (Fig. 6b, $\beta = 20$), the difference in concentration between the tank and the electrode (on average) is very small (approximately 0.5M divided by 20). The reduction in this concentration difference allows the tank and electrode to approach zero state-of-discharge at the end of the first charging cycle (Fig. 6b, zoomed inset plot). Charge utilization is therefore complete in the sense that almost all of the active species are oxidized/reduced in the catholyte/anolyte during the charge cycle. As a result, each charge/discharge cycle starts with tanks having nearly pure state-of-discharge tank at high flow rates, and this effect produces high energy-storage capacity.

Based on our findings of cycling behavior at extreme flow rates we simulated cycling for a range of flow rates to assess its effect on utilization. Two different initial active species concentrations and tank-to-electrode volume ratios with the same amount of active species were simulated (Fig. 7a). For flow rates with $\beta < 2$ the simulations predict near zero utilization. As the flow rate is increased, utilization increases monotonically and about 90% utilization is obtained at $\beta = 20$. At higher flow rates, the porous electrode model predicts the capacities to asymptote close to 100% of theoretical value. Also our simulations show that polarization ($\Delta\phi$) decreases as flow rate increases for $\beta > 2$ (Fig. 7b), and, therefore, energy efficiency losses due to electrochemical processes will be small when operating with high flow rates. For flow rates close to stoichiometric value ($\beta < 2$) the RFB operates like a stationary battery where only the electrolyte in the reactor is utilized towards energy storage. This can be observed in Fig. 7a where utilization approaches a values of $\chi = \dfrac{1}{\alpha + 1}$. Therefore the polarization value decreases as the flow rate approaches zero.

Thus, to facilitate high utilization and efficiency RFBs must be operated at high flow rates, but high flow rate operation necessitates increased pumping pressure that could result in increased pressure-driven crossover of actives and that could require bulky and costly reactor designs to support mechanical loads. Furthermore, the energy required to pump electrolyte could reduce the net energy efficiency of an RFB, including balance-of-plant energy losses. In practice, the choice of operating conditions must be balanced among these factors.

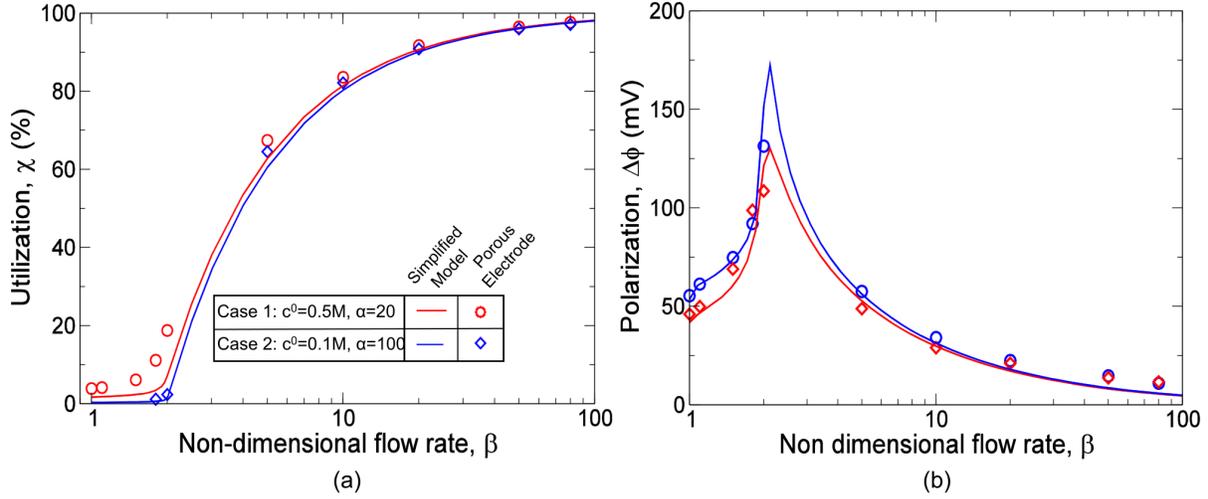

**Figure 7:** Simplified model and porous electrode model predictions of (a) utilization $\chi$ (%) and (b) polarization (mV), with the non-dimensional flow rate for two different cases which have the same theoretical capacity.

*3.3 Utilization and polarization maps from a simplified model*

A simple model is developed to predict the aforementioned behavior of the tank concentration with flow rate and tank size, and from this we estimate the actual capacity that can be obtained after the mixing losses in the tanks during a complete charge/discharge cycle. This model assumes negligible loss in capacity due to kinetic and ohmic polarization, and, therefore, it provides an upper bound on the obtainable capacities that an RFB can achieve with losses due polarization. Figure 1b shows the appropriate control volumes of the cathode half of an RFB during the charging process. By applying species conservation to control volumes around the tank and the entire system the governing equations are:

$$\frac{d(c_{tank}V_{tank})}{dt} = (c_{out} - c_{tank})\dot{V} \text{, and} \qquad (9)$$

$$\frac{d(c_{elec}V_{elec} + c_{tank}V_{tank})}{dt} + \frac{I}{F} = 0, \qquad (10)$$

where $c_{elec}$, $c_{tank}$, and $c_{out}$ are average actives concentration within a given electrode, within its corresponding tank, and exiting the electrode, respectively. To close this set of linear ordinary differential equations (ODEs) we assume that the average electrode concentration is an average of concentrations entering and exiting the reactor (i.e., $c_{elec} = (c_{tank} + c_{out})/2$). We solve these ODEs subject to the initial condition that $c_{tank}(t=0) = c^0$. From this solution the utilization $\chi$ can be determined based by the ratio of actual capacity obtained $Q_{actual}$ relative to $Q_{theory}$, which is equal to a ratio of the actual charge/discharge time $t_{actual}$ to the theoretical charge time $\tau_c$ when galvanostatic conditions are used. After expressing these equations in non-dimensional form, utilization $\chi$ can be determined by a universal function dependent on (1) tank-to-electrode volume ratio $\alpha$ and (2) non-dimensional flow rate $\beta$: $\chi = f(\alpha, \beta)$.

Polarization can also be predicted based on this model, even though it does not include ohmic processes explicitly. This polarization results from irreversible processes within both the tanks and electrodes that are inherent to the chose RFB architecture. Heat generation occurs due to mixing between solution entering the tank and solution stored in the tank, each having different concentrations of active species. Although the local distribution of current within electrodes does not affect the outlet concentration of active species (considering all distributions having the same total current), in the present flow configuration current is biased toward the inlet of the cell because a common cell potential is imposed across the entire electrode, while the state-of-charge of solution varies along its length. Hence, solution will be charged near the inlet with a potential that exceeds its thermodynamically reversible value, and, thus, this effect contributes to polarization. In contrast to ohmic polarization (which scales linearly with current density), this polarization will occur even as applied current approaches zero, and, hence, it represents the minimum energy loss attainable with an RFB configuration of certain tank size $\alpha$ and flow rate $\beta$. We calculate this polarization with the simplified model by assuming that overpotential is negligible at the outlet of both electrodes, in which case cell voltage may be determined as the difference of equilibrium potentials (calculated from Eq. 2) at the outlet of these two electrodes (i.e., $V_{cell} = \phi^+_{eq} - \phi^-_{eq}$). Such analysis shows that, similar to the utilization, polarization $\Delta\phi$ normalized by the "thermal voltage" $\frac{RT}{F}$ is a function only of $\alpha$ and $\beta$ (i.e., $\frac{\Delta\phi F}{RT} = g(\alpha, \beta)$). Hereafter, we refer to polarization produced by the simplified model as the polarization due to inherent irreversibility.

The results of the capacity and polarization predictions of this simplified model were compared to those of the porous electrode model in Fig. 7 and were found to have excellent agreement of over 95%. We also confirm that the effect of tank mixing on utilization that is predicted by the simplified model is consistent with porous electrode simulations even at higher current densities. Specifically, Table 4 compares the charge capacities predicted by the porous electrode model to that of the simplified model for three different current densities of increasing magnitude at low ($\beta = 3$) and high ($\beta = 20$) flow rates respectively. Here, the tank size was increased with current density to achieve a theoretical 5 hour charge/discharge time using the following relationship: $\alpha = i_{app} \tau_c / (c^0 F H \varepsilon) - 1$. Utilization predictions of the simplified model are about 95% in agreement with the predictions of the porous electrode model, indicating that the dependence of utilization on the two dimensionless numbers ($\alpha$ and $\beta$) is true even at higher current densities.

**Table 4**: Comparison of utilization predictions from the porous electrode model and the simplified model at different current densities in low ($\beta=3$) and high ($\beta=20$) non-dimensional flow regimes.

| $\alpha$ | $i_{applied}$ (mA/cm$^2$) | $\beta$ | $\chi_{simulation}$ (%) | $\chi_{simplified}$ (%) | $\Delta\phi$ (mV) |
|---|---|---|---|---|---|
| 128.55 | 1 | 3 | 36.45 | 34.10 | 80.96 |
|  |  | 20 | 91.96 | 90.11 | 20.21 |
| 646.77 | 5 | 3 | 31.89 | 33.48 | 107.14 |
|  |  | 20 | 91.23 | 90.02 | 46.06 |
| 1294.5 | 10 | 3 | 31.34 | 33.41 | 136.2 |
|  |  | 20 | 90.56 | 90.01 | 77.27 |

Now, we examine the variation of polarization with system size, and we show that the polarization due to inherent irreversibility dominates at small system sizes. Figure 8 shows the variation of polarization with the relative total volume of electrolyte with respect to the electrolyte volume in the electrode, $\alpha+1$, obtained from the porous electrode simulations of RFBs. Polarization predictions of the simplified model and a propagating reaction front (PRF) model [52,53] are also shown for comparison with porous electrode model predictions. The PRF model accounts for ohmic resistance from reactions that propagate as a planar front through each electrode, in which case the area-specific resistance $R_{asr}$ is: [52,53]

$$R_{asr} = \left( \frac{H}{2\kappa_{eff}} + \frac{H_{sep}}{\kappa_0 \varepsilon_{sep}/\tau_{sep}} + \frac{H}{2\kappa_{eff}} \right). \quad (12)$$

Here, ionic conductivity values used with this model were chosen consistent with porous electrode simulations. Polarization due to the PRF can be determined as the product of $R_{asr}$ and the applied current density $i_{app}$. The results are shown for two different concentrations in a high flow rate regime of operation $\beta=20$, and current density is adjusted to obtain the same theoretical charge/discharge time of 5 hours among all different cases, such that current density increases linearly with $\alpha$. The polarization due to inherent irreversibilty asymptotes to a particular value of 10mV, while for sufficiently large tanks (e.g., $\alpha=1000$) the polarization due to area-specific resistance is 100mV, exceeding that due to inherent irreversibility by an order of magnitude because of high current densities needed for large tanks. Thus it can be concluded that for smaller tank sizes the polarization due to the inherent irreversibilities is the dominant source of polarization, whereas for larger tank sizes the polarization due to the area-specific resistance dominates.

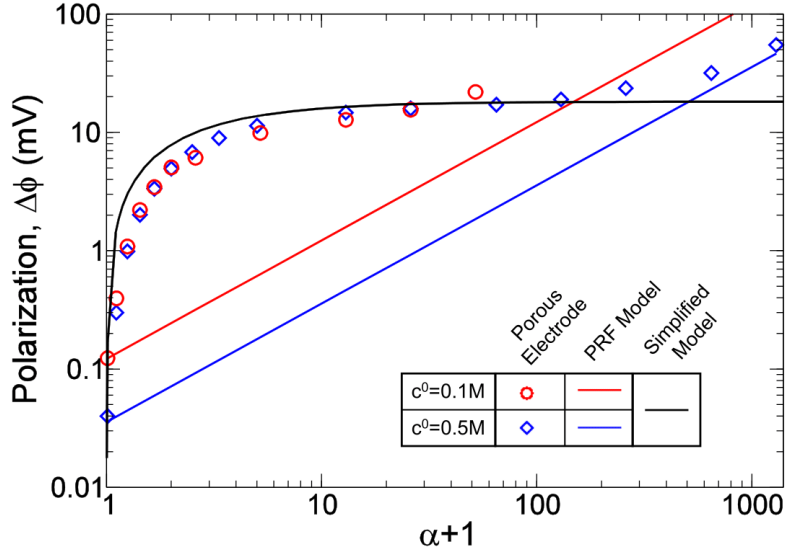

**Figure 8:** Variation of polarization with the ratio of total volume of electrolyte in the RFB to the volume of electrolyte in the electrode. The RFB is operated at a high non-dimensional flow rate $\beta=20$ for two different active-species concentrations. Polarization predictions from the propagating reaction front model and the simplified model are also shown for comparison.

Since the simplified model predicts utilization and polarization levels similar to the porous electrode model at low current densities, it can be used as an upper-bound estimate of utilization and a lower-bound estimate of polarization for RFBs having various tank sizes and using various flow rates. Figure 9 shows the variation of (a) utilization and (b) non-dimensional polarization with flow rate for various tank-to-electrode volume ratios, as predicted by the simplified model. The charge capacity increases and polarization due to inherent irreversibilities decreases with increasing flow rate. Although the percentage capacity is higher for smaller tank sizes, the base theoretical capacity for such systems is small. For larger tanks, $\alpha>20$, the degree of change in utilization and polarization with tank size is small indicating that the utilization and polarization are governed by flow rate alone $\chi=f(\alpha,\beta)\approx f^*(\beta)$ and $\Delta\phi=g(\alpha,\beta)\approx g^*(\beta)$ when $\alpha>20$. For flow rates near stoichiometric flow velocity ($\beta=1$), the capacity that can be obtained is very low. The charge utilization increases with increasing flow rate and asymptotes to near 100% capacity for very high flow rates ($\beta=100$).

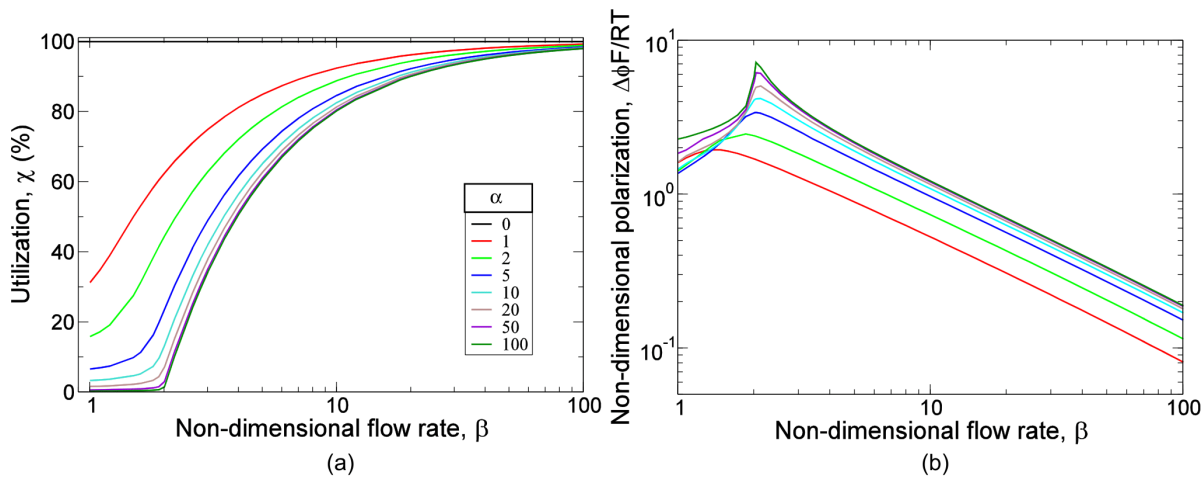

**Figure 9:** Simplified model predictions of (a) utilization capacity and (b) non-dimensional polarization as a function of the dimensionless flow rate $\beta$ for various tank-to-electrode volume ratios $\alpha$.

## 4. Conclusions

We have developed mechanistic understanding of how flow rate affects RFB utilization with the aid of a model coupling reactor and tank transport processes. While much of RFB research focuses on developing redox-active materials and reactor designs, the present results suggest that mixing within tanks can be a significant source of capacity loss. These losses in capacity were distinguished from those resulting from resistive polarization effects by coupling the reaction distribution within porous electrodes to well-mixed tanks. While at high flow rates mixing effects are negligible, low flow-rate capacity losses are dominated by the effects of mixing within tanks. Based on a simplified model including these effects we identified two dimensionless numbers, the tank-to-electrode volume ratio and dimensionless flow rate (ratio of actual flow rate to the so called stoichiometric flow rate), which affect the RFB's charge utilization. The simplified model is about 95% accurate to the predictions of the continuum model and therefore can be used as a tool to predict the maximum capacity one can expect from the system at a particular operating condition. The RFB must be operated at least twenty times the stoichiometric flow rate in order to extract 90% of the theoretical capacity. The mechanism which governs the polarization in the reactor is also explored by plotting the spatial variation of the pore scale reaction current density which is localized at low flow rates (leading to higher polarization) and more uniformly spread at high flow rates (leading to lower polarization). The spatial variation of current densities along with thermodynamic losses due to mixing in the tanks is referred to as inherent irreversibility of the RFBs and the polarization due to this inherent irreversibility dominates the polarization due to area-specific resistance of the reactor at relatively smaller tanks ($\alpha < 100$). One can expect high utilization and low polarization losses for high flow rates ($\beta > 20$). But this comes with a trade-off of high pumping pressure, which can affect system stability and increase crossover losses. Though not captured by this model, the crossover of active species due to pressure differences and concentration gradients across the membrane could also play a role in determining capacity.

The modeling results presented here highlight the importance of mixing processes within RFB tanks. The particular model presented here assumes perfectly mixed tanks. In reality turbulence, diffusion, and active mixing processes within tanks will affect the degree of mixing. However, this idealized mixing condition sets a benchmark for performance and avoiding/minimizing mixing with appropriate tank design could lead to a better performing RFB. Also, the assumption of convection domination over diffusion (high Peclet number) can be contested depending on the electrochemical motif of interest. Also, the present results motivate the simultaneous optimization of capacity, polarization, and pumping pressure. The relative importance of each of these factors will depend on the scale of demonstration (e.g., for lab or grid demonstrations).

**Acknowledgments**

This work was supported as a part of the Joint Center for Energy Storage Research, an Energy Innovation Hub funded by the U.S. Department of Energy, Office of Science, Basic Energy Sciences.

**List of Symbols**

| | |
|---|---|
| $a$ | volumetric surface, 1/m |
| $\alpha$ | tank-to-electrode volume ratio |
| $\beta$ | ratio of flow rate to stoichiometric flow rate |
| $c^0$ | initial concentration, M |
| $c_R$ | concentration of reduced species, M |
| $E^0$ | equilibrium potential at 50% state of charge, V |
| $\varepsilon$ | porosity |
| $D_i^{eff}$ | effective diffusion coefficient, m$^2$/s |
| $F$ | faraday's constant, C/mol |
| $\gamma$ | tortuosity scaling-exponent |
| $H$ | electrode thickness, m |
| $\eta$ | overpotential, V |
| $i_{app}$ | applied current density, A/m$^2$ |
| $i_0$ | exchange current density, A/m$^2$ |
| $i_n$ | reaction current density, A/m$^2$ |
| $\kappa_0$ | bulk ionic conductivity, S/m |
| $K$ | permeability, m |
| $L$ | length of the electrode unit cell, m |
| $n_e$ | number of electron transfer in redox reaction |
| $\mu$ | viscosity, Pa.s |
| $v_s$ | solid volume fraction |
| $\vec{N}$ | species flux, mol/m$^2$-s |
| $Pe$ | peclet number |
| $\phi_e$ | solution-phase potential, V |
| $\phi_{eq}$ | equilibrium potential, V |
| $\phi_s$ | solid-phase potential, V |
| $\psi$ | state of discharge |
| $Q$ | theoretical capacity, C |
| $R$ | universal gas constant, J/mol-K |
| $\sigma_s$ | effective electronic conductivity, S/m |
| $\tau$ | tortuosity |
| $\tau_c$ | theoretical charge time, s |
| $\tau_{stoich}$ | stoichiometric residence time, s |
| $t$ | time, s |
| $T$ | temperature, K |
| $\vec{u}$ | velocity field, m/s |
| $\dot{V}$ | volumetric flow rate, m$^3$/s |
| $z_i$ | valence number |